\documentclass[preprint]{revtex4}
\usepackage{graphicx}
\renewcommand{\vec}[1]{\mbox{\boldmath $#1$}}
\begin{document}
\title{Free Expansion of a Weakly-interacting Dipolar Fermi Gas}
\author{Takushi Nishimura}
\email[]{nishimura.takushi@ocha.ac.jp}
\affiliation{Division of Advanced Sciences, Ochadai Academic Production, Ochanomizu University, Otsuka, Bunkyo, Tokyo 112-8610, Japan}
\author{Tomoyuki Maruyama}
\email[]{maruyama.tomoyuki@nihon-u.ac.jp}
\affiliation{College of Bioresource Sciences, Nihon University, Fujisawa 252-8510, Japan}
\affiliation{Advanced Science Research Center, Japan Atomic Energy Agency, Tokai 319-1195, Japan}
\begin{abstract}
We theoretically investigate 
a polarized dipolar Fermi gas 
in free expansion. 
The inter-particle dipolar interaction 
deforms phase-space distribution 
in trap and also in the expansion. 
We exactly predict 
the minimal quadrupole deformation in the expansion 
for the high-temperature Maxwell-Boltzmann 
and zero-temperature Thomas-Fermi gases 
in the Hartree-Fock and Landau-Vlasov approaches. 
In conclusion, 
we provide a proper approach 
to develop the time-of-flight method 
for the weakly-interacting dipolar Fermi gas 
and also reveal a scaling law 
associated with the Liouville's theorem 
in the long-time behaviors of the both gases. 
\end{abstract} 
%
%
\maketitle
%
%
Recent development of trapping and manipulating techniques of atoms 
enables to realize ultracold dipolar gases of atoms and molecules 
in several years. 
In fact, 
dipolar Bose gases are realized 
in ${}^{52}$Cr atoms 
with magnetic dipole moments~\cite{dipolar-B-ex-1,dipolar-B-ex-2,dipolar-B-ex-3}, 
and dipolar Fermi gases are also realized 
in heteronuclear ${}^{40}$K-${}^{87}$Rb molecules 
with electric dipole moments~\cite{dipolar-F-ex-1,dipolar-F-ex-2,dipolar-F-ex-3}. 
The dipolar gases offer 
a great infrastructure for quantum many-body physics 
as new types of actual quantum systems. 

The dipolar gases have 
a tensor-type dipolar interaction~\cite{Landau}, 
which depends on not only inter-particle relative position 
but also directions of the dipole moments. 
When we suppose the dipolar fermions 
perfectly polarized in a strong external electric field 
along the $z$-axis of the Cartesian axes 
as in the actual experiments~\cite{dipolar-F-ex-1,dipolar-F-ex-2,dipolar-F-ex-3}, 
the interaction becomes 
\begin{equation}
v(\vec{r}) 
= G_d \frac{1 - 3 r_z^2 / r^2}{r^3} 
\label{Eq-4}
\end{equation}
with $\vec{r} \equiv \{ r_x, r_y, r_z \}$, 
$r \equiv \left| \vec{r} \right|$, 
and a positive coupling constant $G_d \propto d^2$ 
given later with some scaling units, 
where the electric dipolar moment $d$ 
is experimentally manipulated in 
$d \sim 10^{-11 \sim 0} e a_{\text{B}}$ 
by choosing molecular internal states 
and electric field strength~\cite{dipolar-F-ex-1,dipolar-F-ex-2,dipolar-F-ex-3}. 

As an important feature, 
the dipolar Fermi gas is purely dominated 
by the long-range tensor force 
because of vanishment of short-range interactions 
owing to the Pauli blocking effect. 
The axisymmetric interaction in eq.~(\ref{Eq-4}) 
produces anisotropic quantum correlations 
and density deformation 
and also makes instability 
in strong interaction regime~\cite{Baranov, Goral, Miyakawa, Sogo, He}. 
Furthermore, differently from the short-range interactions, 
the long-range dipolar interaction deforms 
phase-space distribution, 
which indicates important information 
for quantum many-body effects in the dipolar gas~\cite{Miyakawa, Sogo}. 

Free expansion physics is quite important for the dipolar gas. 
That is because the expansion is often used 
to observe the momentum distribution of the trapped gas 
in the time-of-flight (TOF) method; 
and the anisotropic momentum deformation directly reflects 
the phase-space deformation and interaction effects 
in the dipolar gas. 
In fact, the dipolar Bose gases are firstly observed 
in the expansion~\cite{dipolar-B-ex-1,dipolar-B-ex-2,dipolar-B-ex-3}. 
The dipolar interaction may change 
the momentum distribution not only in trap 
but also in the expansion, 
then we need some corrections 
to apply the TOF method to obtain the initial momentum distribution 
differently from the ballistic expansion 
with a conserved momentum distribution. 
To obtain the corrections, 
some people develop the quadrupole scaling method~\cite{He, Sogo}; 
however no work has been done 
to demonstrate exact time-evolution of the expanding Fermi gas 
and confirm validity of the scaling method in the expansion. 
The expansion problem is one of the most important current issues 
in theoretical physics on the dipolar gas. 

The aim of this work is to reveal the expansion problem, 
i.e. relationship between the initial and final momentum deformations, 
in weak interaction regime. 
Here we exactly evaluate 
time-evolution of the minimal quadrupole deformation 
of the expanding gas 
by developing another scaling ansatz, 
which reproduces the Liouville's theorem 
and perturbation theory from the ballistic expansion. 
This work must provide 
a proper correction for the TOF method 
to detect the quantum many-body effects on the dipolar gas. 

Let us define 
the Hamiltonian of the dipolar gas as 
\begin{equation}
H 
= \int d{\vec{r}}~ \psi^{\dagger}(\vec{r}) \left[ - \frac{1}{2} \vec{\nabla}_{\vec{r}}^2 + v_{\text{T}}(\vec{r}) \right] \psi(\vec{r}) 
+ H_{\text{I}} 
\label{Eq-2}
\end{equation}
with the fermion field operator $\psi$ and 
interaction part 
\begin{equation}
H_{\text{I}} 
\equiv \int d{\vec{r}} \int d{\vec{r}^{\prime}} \psi^{\dagger}(\vec{r}) \psi^{\dagger}(\vec{r}^{\prime}) v(\vec{r} - \vec{r}^{\prime}) \psi(\vec{r}^{\prime}) \psi(\vec{r}), 
\label{Eq-3}
\end{equation}
where $v$ is the dipolar interaction in eq.~(\ref{Eq-4}). 
The trap term is given as 
$v_{\text{T}}(\vec{r}) = \Theta(-t) V_{\text{T}}(\vec{r})$ 
with the step function $\Theta(x)$ 
and trap potential $V_{\text{T}}(\vec{r})$ before $t = 0$, 
when we leave the gas. 
To simplify the notations, 
we choose units for the reduced Planck constant and mass 
as $\hbar = 1$ and $m = 1$ 
and omit the time parameter $t$, 
e.g. $\psi(\vec{r}, t) \equiv \psi(\vec{r})$. 

To treat $H_{\text{I}}$ in eq.~(\ref{Eq-2}), 
we apply 
the time-dependent Hartree-Fock approximation (TDHFA)~\cite{RingSchuck}, 
in which 
the two-body interaction $H_{\text{I}}$ 
is rewritten into one-body interactions 
with self-consistent mean-fields, 
the direct (Hartree) and exchange (Fock) terms, 
reflecting the many-body effects as mean-values. 
In TDHFA, 
the field operator 
satisfies an equation of motion, 
\begin{equation}
i \frac{d \psi(\vec{r})}{d t} 
= - \frac{1}{2} \vec{\nabla}_{\vec{r}}^2 \psi(\vec{r})
+ \int d{\vec{r}^\prime} U(\vec{r}, \vec{r}^\prime) \psi(\vec{r}^\prime), 
\label{Eq-6}
\end{equation}
with the self-consistent non-local self-energy 
$U(\vec{r}, \vec{r}^\prime)$ 
given later in eq.~(\ref{Eq-16}). 
Then 
the Wigner function~\cite{Wigner}, 
\begin{equation}
f(\vec{r}, \vec{p}) 
\equiv \int d{\vec{s}}~ e^{- i \vec{p} \cdot \vec{s}} 
{\bf n}\left( \vec{r} - \frac{\vec{s}}{2}, \vec{r} + \frac{\vec{s}}{2} \right) 
\label{Eq-12}
\end{equation}
with the density matrix 
${\bf n}(\vec{r}, \vec{r}^\prime) \equiv \bigl< \psi^\dagger(\vec{r}^\prime) \psi(\vec{r}) \bigr>$, 
obeys 
\begin{equation}
\frac{d f}{d t}
= 
- \vec{p} \cdot \vec{\nabla}_{\vec{r}} f 
+ \frac{2}{\hbar} 
\sin\left[ \frac{\vec{\nabla}_{\vec{p}}^{f} \cdot \vec{\nabla}_{\vec{r}}^{u} - \vec{\nabla}_{\vec{p}}^{u} \cdot \vec{\nabla}_{\vec{r}}^{f}}{2 \hbar^{-1}} \right] 
u f 
\label{Eq-15}
\end{equation}
with the Wigner transformation of 
$U(\vec{r}, \vec{r}^\prime)$, 
\begin{eqnarray}
u(\vec{r}, \vec{p}) 
&\equiv& \int d{\vec{s}}~ 
e^{- i \vec{p} \cdot \vec{s}}  U\left( \vec{r} + \frac{\vec{s}}{2}, \vec{r} - \frac{\vec{s}}{2} \right) 
\nonumber \\ 
&=& v_{\text{T}}(\vec{r}) + u_{\text{H}}(\vec{r}) + u_{\text{F}}(\vec{r}, \vec{p}), 
\label{Eq-16}
\end{eqnarray}
where $\vec{\nabla}^{f}$ and $\vec{\nabla}^{u}$ act 
on $f$ and $u$, respectively. 
Here we introduce 
the Hartree part 
\begin{equation}
u_{\text{H}}(\vec{r}) 
\equiv \int d\vec{r}^\prime v(\vec{r} - \vec{r}^\prime) n(\vec{r}^\prime) 
\label{Eq-10}
\end{equation}
with the number density 
$n(\vec{r}) \equiv {\bf n}(\vec{r}, \vec{r})$ 
and 
Fock part 
\begin{equation}
u_{\text{F}}(\vec{r}, \vec{p})
\equiv - \int \frac{d{\vec{q}}}{(2 \pi)^3} V(\vec{p} - \vec{q}) f(\vec{r}, \vec{q}) 
\label{Eq-17}
\end{equation}
with the Fourier transformation of 
$v(\vec{r})$ in eq.~(\ref{Eq-4}), 
\begin{equation}
V(\vec{p}) 
\equiv \int d{\vec{r}}~ e^{- i \vec{p} \cdot \vec{r}} v(\vec{r})
= - \frac{4 \pi}{3} G_{d} \frac{p^2 - 3 p_{z}^2}{p^2}, 
\label{Eq-18}
\end{equation}
where $p \equiv \left| \vec{p} \right|$. 
We explicitly write $\hbar$ ($= 1$) 
only in eq.~(\ref{Eq-15}) 
for the later explanation. 

To treat the quantum dynamics in eq.~(\ref{Eq-15}), 
we use the semi-classical Landau-Vlasov (LV) approach~\cite{Vlasov}; 
i.e. expanding the trigonometric function in eq.~(\ref{Eq-15}) 
up to the first order of $\hbar$. 
Then we obtain the LV equation, 
\begin{equation}
\frac{d f}{d t} 
= - \left( \vec{p} + \vec{\nabla}_{\vec{p}} u \right) \cdot \left( \vec{\nabla}_{\vec{r}} f \right) 
+ \left( \vec{\nabla}_{\vec{r}} u \right) \cdot \left( \vec{\nabla}_{\vec{p}} f \right). 
\label{Eq-19}
\end{equation}
As a conclusion of the formulation, 
we study the expansion 
by solving the LV equation, eq.~(\ref{Eq-19}), 
with some initial conditions, 
$f(\vec{r}, \vec{p}; t = 0) = f_{0}(\vec{r}, \vec{p})$. 

Here we emphasize that 
a feature of the finite-range interaction 
(e.g. eq.~(\ref{Eq-4})) 
must be in contribution of the Fock part, 
i.e. the momentum dependence of $u$ as shown in eq.~(\ref{Eq-16}), 
so that $\vec{\nabla}_{\vec{p}} u$ in eq.~(\ref{Eq-19}) 
may give notable contribution in the dipolar gas 
and should not be omitted. 

In this work, 
we focus on the interaction effects 
on the time-dependent momentum distribution, 
$\rho(\vec{p}) = \int d{\vec{r}} f(\vec{r}, \vec{p})$, 
which corresponds to the density distribution 
at long-time limit as 
\begin{equation}
n(\vec{r}) 
= \frac{1}{(2 \pi t)^3} \rho\left( \frac{\vec{r}}{t} \right) + O(t^{-1}). 
\label{Eq-23}
\end{equation}
That is because we can pragmatically neglect the interaction 
after a cut-off time $t_{\text{c}}$ 
($\to \infty$ in this theoretical work) 
~\cite{footnote-1}
, 
when the Wigner function in eq.~(\ref{Eq-19}) 
is regarded as the ballistic solution, 
\begin{equation}
f(\vec{r}, \vec{p}) 
= f_{0}(\vec{\xi}, \vec{p}), 
\label{Eq-19-0}
\end{equation}
with the Galilei transformation, 
$\vec{r} \to \vec{\xi} \equiv \vec{r} - \vec{p} t$; 
then 
the density distribution at $t > t_{\text{c}}$ becomes 
\begin{eqnarray}
n(\vec{r}) 
&=& \int \frac{d{\vec{p}}}{(2 \pi)^3} f_{\text{c}}(\vec{r} - \vec{p} (t - t_{\text{c}}), \vec{p}) 
\nonumber \\ 
&=& \int \frac{d{\vec{\xi}}}{(2 \pi t)^3} f_{\text{c}}\left( \vec{\xi} + \vec{p} t_{\text{c}}, \frac{\vec{r}}{t} \right) + O(t^{-1}) 
\label{Eq-24}
\end{eqnarray}
with the Wigner function $f_{\text{c}}(\vec{r}, \vec{p})$ 
at $t = t_{\text{c}}$, 
where eq.~(\ref{Eq-24}) agrees with eq.~(\ref{Eq-23}) 
owing to the conservation of the momentum distribution 
in the ballistic expansion at $t > t_{\text{c}}$. 
Thus the interaction effects appear 
in the time-evolution before $t_{\text{c}}$, 
and one can obtain the final momentum distribution 
from the density distribution at long-time limit 
as shown in eq.~(\ref{Eq-23}). 

Before calculation of the expansion dynamics, 
we firstly consider the ballistic expansion; 
i.e. the interaction is neglected 
only in the expansion process 
except in the initial states. 
In the ballistic expansion, 
the LV equation, eq.~(\ref{Eq-19}), becomes 
\begin{equation}
\frac{d f}{d t} 
= - \vec{p} \cdot \vec{\nabla}_{\vec{r}} f. 
\label{Eq-19-1}
\end{equation}
Its solution exactly agrees with 
the ballistic solution in eq.~(\ref{Eq-19-0}). 
When we assume the initial Wigner function 
as some function of a parameter 
~\cite{footnote-2}
\begin{equation}
z(\vec{r}, \vec{p}) 
= \sum_{l = x, y, z} \bigl( \gamma_{l}^{2} p_{l}^{2} + \omega_{l}^{2} r_{l}^{2} \bigr), 
\label{Eq-19-2}
\end{equation}
the dynamics is determined only by a time-dependent parameter 
$\bar{z} \equiv z(\vec{\xi}, \vec{p})$, 
which becomes 
$\bar{z} = \omega_{l}^{2} p_{l}^{2} t^{2} - 2 \omega_{l}^{2} r_{l} p_{l} t + O(t^{0})$ 
at the long-time limit. 

In the previous approach~\cite{Sogo}, 
the dynamical parameter is approximated as 
$\bar{z} \approx z(\vec{r}^{\prime}, \vec{p}^{\prime})$ 
with $r_{l}^{\prime} \equiv b_{l}^{-1} r_{l}$ 
and $p_{l}^{\prime} \equiv b_{l} p_{l} - \dot{b}_{l} r_{l}$, 
where $b_{l}$ obeys 
$\ddot{b}_{l} = D_{l}^2 b_{l}^{-3}$ 
with 
$D_{l} \equiv \sqrt{\bigl< p_{l}^{2} \bigr>(t = 0) / \bigl< r_{l}^{2} \bigr>(t = 0)}$ 
and 
$b_{l}(t = 0) = 1$ 
and becomes 
$b_{l} = D_{l} t + 1 / (2 D_{l} t) + O(t^{-3})$ 
at the long-time limit, after all. 
Here the cap-dot represents the Newton's notation for time derivative. 
Owing to the quadratic Wigner function, 
$D_{l} = \omega_{l} / \gamma_{l}$, 
and 
$\bar{z} \approx \omega_{l}^{2} p_{l}^{2} t^{2} - 2 \omega_{l}^{2} r_{l} p_{l} t + O(t^{0})$ 
in the previous ansatz. 
This result agrees with that of the exact solution. 
However this agreement is a specific property 
of the quadratic Wigner function. 
Our ansatz in this work 
generally satisfies the ballistic limit 
beyond the quadratic case 
and gives the exact solution for the minimal variation 
from the ballistic expansion 
as described below. 

We now consider 
the quadrupole deformation of the momentum distribution 
in the interacting non-ballistic expansion. 
The quadrupole deformation is 
the most important deformation 
owing to the axisymmetric interaction in eq.~(\ref{Eq-4}) 
and reflects not only 
the initial deformation in trap 
but also the additional deformation 
in the expansion. 
To describe them, 
we introduce a time-dependent index parameter: 
$\lambda_{p} \equiv \ln( \mathcal{T}_{z} / \mathcal{T}_{0} )$ 
with 
$\mathcal{T}_{z} \equiv \bigl< p_z^2 \bigr>$ 
and 
$\mathcal{T}_{0} \equiv ( \bigl< p_x^2 \bigr> \bigl< p_y^2 \bigr> \bigl< p_z^2 \bigr> )^{1/3}$; 
then our interest is just focused on relationship 
between the initial conditions and 
$\lambda_{p}(t \to \infty)$. 
Eq.~(\ref{Eq-19}) leads 
\begin{equation}
\frac{d \lambda_{p}}{d t} 
= - \int d{\vec{\xi}} \int \frac{d{\vec{p}}}{(2 \pi)^3} \frac{2 p_z}{\mathcal{T}_{z}} \frac{\partial u(\vec{\xi} + \vec{p} t, \vec{p})}{\partial \xi_z} \tilde{f}(\vec{\xi}, \vec{p}), 
\label{Eq-26}
\end{equation}
where we define 
$\tilde{f}(\vec{\xi}, \vec{p}) \equiv f(\vec{\xi} + \vec{p} t, \vec{p})$ 
and use the partial integral. 
According to the definition in eq.~(\ref{Eq-16}), 
$u(\vec{\xi} + \vec{p} t, \vec{p})$ in eq.~(\ref{Eq-26}) 
can be written as 
\begin{equation}
u(\vec{\xi} + \vec{p} t, \vec{p}) 
= \int d{\vec{\eta}} \int \frac{d{\vec{q}}}{(2 \pi)^3} \bar{v}(\vec{\xi} - \vec{\eta}, (\vec{p} - \vec{q}) t) \tilde{f}(\vec{\eta}, \vec{q}), 
\label{Eq-30}
\end{equation}
where we introduce 
\begin{equation}
\bar{v}(\vec{x}, \vec{a} t) 
\equiv v(\vec{x} + \vec{a} t) - V(\vec{a}) \delta(\vec{x} + \vec{a} t) 
\label{Eq-28}
\end{equation}
with $\vec{q} = \vec{a} + \vec{p}$ and $\vec{\eta} = \vec{x} + \vec{\xi}$. 
By substituting eq.~(\ref{Eq-30}) into eq.~(\ref{Eq-26}) 
and using exchange symmetry on 
$\{ \vec{\xi}, \vec{p} \} \leftrightarrow \{ \vec{\eta}, \vec{q} \}$, 
we obtain 
\begin{equation}
\frac{d \lambda_{p}}{d t} 
= \int d{\vec{x}} \int \frac{d{\vec{a}}}{(2 \pi)^3} \frac{a_z}{\mathcal{T}_{z}} \bar{v}(\vec{x}, \vec{a} t) \frac{\partial F(\vec{x}, \vec{a})}{\partial x_z} 
\label{Eq-27}
\end{equation}
with 
\begin{equation}
F(\vec{x}, \vec{a}) 
\equiv \int d{\vec{\xi}} \int \frac{d{\vec{p}}}{(2 \pi)^3} \tilde{f}(\vec{\xi}, \vec{p}) \tilde{f}(\vec{\xi} + \vec{x}, \vec{p} + \vec{a}). 
\label{Eq-29}
\end{equation}

We here assume small variation from the ballistic solution 
in the phase-space distribution 
and introduce an ansatz for the Wigner function as 
\begin{equation}
\tilde{f}(\vec{\xi}, \vec{p})
= f_{0}(\tilde{\vec{\xi}}, \tilde{\vec{p}}) 
\label{Eq-34}
\end{equation}
with 
\begin{eqnarray}
\tilde{p}_{x, y} \equiv e^{\lambda / 4} \bigl( p_{x, y} + \frac{\dot{\lambda}}{4} \xi_{x, y} \bigr), 
&& 
\tilde{p}_{z} \equiv e^{- \lambda / 2} \bigl( p_{z} - \frac{\dot{\lambda}}{2} \xi_{z} \bigr), 
\nonumber \\ 
\tilde{\xi}_{x, y} \equiv e^{- \lambda / 4} \xi_{x,y}, 
&& 
\tilde{\xi}_{z} \equiv e^{\lambda / 2} \xi_{z}, 
\label{Eq-35}
\end{eqnarray}
where $\lambda$ indicates 
the additional deformation in the expansion, 
$\lambda_{p} \simeq \lambda_{0} + \lambda$, 
with $\lambda(t = 0) = 0$. 
At the ballistic limit, 
$\lambda = 0$, 
and it always reproduces the ballistic solution. 
Furthermore 
this ansatz exactly satisfies 
the Liouville's theorem 
corresponding to 
the current equation in the microscopic theory~\cite{TMTN,TNTM}. 

In this Letter, 
we consider two general initial conditions: 
the high-temperature Maxwell-Boltzmann (MB) 
and zero-temperature Thomas-Fermi (TF) gases 
in cylindrical harmonic oscillator traps 
with the small phase-space quadrupole deformation 
owing to the weak interaction. 
The initial Wigner function becomes 
\begin{equation}
f_{0}(\vec{r}, \vec{p}) 
= e^{- ( R_c + R_z + P_c + P_z ) / 2} 
\label{Eq-31}
\end{equation}
for the MB gas and 
\begin{equation}
f_{0}(\vec{r}, \vec{p}) 
= \Theta\left[ 1 - \frac{R_c + R_z + P_c + P_z}{2} \right] 
\label{Eq-33}
\end{equation}
for the TF gas with 
\begin{eqnarray}
R_c \equiv e^{(\Lambda_{0} - \lambda_{0}) / 2} (r_{x}^2 + r_{y}^2), 
 && 
R_z \equiv e^{- (\Lambda_{0} - \lambda_{0})} r_{z}^2, 
\nonumber \\ 
P_c \equiv e^{\lambda_{0} / 2} (p_{x}^2 + p_{y}^2), 
 && 
P_z \equiv e^{- \lambda_{0}} p_{z}^2, 
\label{Eq-32}
\end{eqnarray}
where $\lambda_{0} \equiv \lambda_{p}(t = 0)$ 
indicates the initial momentum deformation 
~\cite{footnote-3}
, 
and $\Lambda_{0}$ indicates 
sum of the initial density and momentum deformations. 
In addition, 
we also determine $G_d$ in eq.~(\ref{Eq-4}) as 
$G_d = ( k_{\text{B}} T / ( \hbar \omega ) )^{- 5 / 2} N ( a_{\text{B}} / a_{\text{ho}} ) ( m / m_{e} ) ( d / (e a_{\text{B}}) )^2$ for the MB gas 
and 
$G_d = ( 6 N )^{1 / 6} ( a_{\text{B}} / a_{\text{ho}} ) ( m / m_{e} ) ( d / (e a_{\text{B}}) )^2$ 
for the TF gas 
~\cite{footnote-4}
with the Boltzmann constant $k_{\text{B}}$, 
temperature $T$, 
particle number $N$, 
Bohr radius $a_{\text{B}}$, 
electron mass $m_{e}$, 
elementary electric charge $e$, 
trap frequency 
$\omega \equiv ( \omega_{x} \omega_{y} \omega_{z} )^{1 / 3}$ 
($= 1$ in this Letter), 
and oscillator length 
$a_{\text{ho}} \equiv \sqrt{\hbar / ( m \omega )}$. 
The above formulations reproduce general quadratic descriptions 
of the MB and TF gases with rescaling 
~\cite{footnote-5}
. 

Here we should comment that 
the interaction effect on $\Lambda_{0}$ in eq.~(\ref{Eq-32})
is always cancelled out 
in the quadrupole deformation 
owing to conservation of phase-space volume 
in the Liouville's theorem. 
Thus $\Lambda_{0}$ corresponds to 
the density deformation of a non-interacting gas 
in the initial trap potential, 
i.e. 
$\Lambda_{0} = \ln{( \omega_{x} / \omega_{z} )^{4 / 3}} = \ln{( \omega_{y} / \omega_{z} )^{4 / 3}}$. 

According to eqs.~(\ref{Eq-34}), (\ref{Eq-31}), and (\ref{Eq-33}), 
$F(\vec{x}, \vec{a})$ in eq.~(\ref{Eq-29}) becomes 
\begin{equation}
F(\vec{x}, \vec{a}) 
= \frac{1}{8} e^{- s^2 / 4} 
\label{Eq-39}
\end{equation}
for the MB gas and 
\begin{eqnarray}
F(\vec{x}, \vec{a}) 
&=& \frac{\Theta\left( 8 - s^2 \right)}{3 \pi} \bigl[ 
\arcsin\bigl( \frac{\sqrt{8 - s^2}}{\sqrt{8}} \bigr) 
\nonumber \\ 
&& - \frac{s \sqrt{8 - s^2}}{960} ( s^4 - 26 s^2 + 264 ) \bigr] 
\label{Eq-40}
\end{eqnarray}
for the TF gas, where 
\begin{equation}
s 
\equiv \sqrt{ \sum_{j = x, y} A_{j} + A_{z} + \sum_{j = x, y} X_{j} + X_{z} } 
\label{Eq-37}
\end{equation}
with 
\begin{eqnarray}
A_{j} \equiv e^{\lambda_{p} / 2} \bigl( a_{i} + \frac{\dot{\lambda}}{4} x_{i} \bigr)^2, 
 && 
A_{z} \equiv e^{- \lambda_{p}} \bigl( a_{z} - \frac{\dot{\lambda}}{2} x_{z} \bigr)^2, 
\nonumber \\ 
X_{j} \equiv e^{(\Lambda_{0} - \lambda_{p}) / 2} x_{j}^2, 
 && 
X_{z} \equiv e^{- (\Lambda_{0} - \lambda_{p})} x_{z}^2. 
\label{Eq-38}
\end{eqnarray}
Substituting eq.~(\ref{Eq-39}) or (\ref{Eq-40}) 
into eq.~(\ref{Eq-27}) 
and expanding up to the first orders 
of $\Lambda_{0}$, $\lambda_{p}$, and $\dot{\lambda}$, 
we obtain a time-evolution equation for the minimal deformation: 
\begin{eqnarray}
\dot{\lambda} 
&=& G_d \sum_{I = 0}^{\infty} \sum_{J = 0}^{\infty} \sum_{K = 0}^{\infty} 
\Gamma_{IJK} \Lambda_{0}^{I}~ \lambda_{p}^{J}~ \dot{\lambda}^{K} 
\nonumber \\ 
&\simeq& G_d \left( \Gamma_{000} + \Gamma_{100} \Lambda_{0} + \Gamma_{010} \lambda_{p} + \Gamma_{001} \dot{\lambda} \right) 
\label{Eq-41}
\end{eqnarray}
with the time-dependent coefficients $\Gamma_{I J K}(t)$ 
depending on the initial conditions. 

After the calculation, 
we obtain $\Gamma_{000} = \Gamma_{010} = \Gamma_{001} = 0$ 
and 
\begin{equation}
\dot{\lambda} 
\simeq G_d \Lambda_{0} \Gamma_{100} 
\label{Eq-41-2}
\end{equation}
in the both gases. 
It reveals that 
the additional deformation $\lambda$ in the expansion 
depends only on a conserved quantity $G_d \Lambda_0 \equiv \mathfrak{S}$ 
and exhibits a scaling behavior, 
\begin{equation}
\lambda(t \to \infty) 
\simeq \gamma_{100} G_d \Lambda_0 
= \gamma_{100} \mathfrak{S}, 
\label{Eq-44}
\end{equation}
at long-time limit 
with $\gamma_{100} \equiv \int_{0}^{\infty} dt~ \Gamma_{100}$. 
The scaling behavior is due to 
cancellation of the Hartree and Fock parts 
determined by the time-developing phase-space distribution 
with the quadrupole deformation. 
According to the Liouville's theorem, 
the incompressible quadrupole deformation in the momentum space 
gives strong constraint for the spacial motion 
as shown in the description of the deformation ansatz, 
and this constraint provides a source 
of the cancellation and scaling behavior. 

Here it should be noted that 
the minimal deformation in eq.~(\ref{Eq-41-2}) 
exactly agrees with that in the first order perturbation theory, 
which can be obtained 
by replacing $F(\vec{x}, \vec{a})$ defined in eq.~(\ref{Eq-27}) 
with that of the ballistic solution, 
i.e. substituting 
$\lambda = 0$ and $\dot{\lambda} = 0$ 
in eqs.~(\ref{Eq-39}) and (\ref{Eq-40}). 
This agreement is obviously due to the cancellation 
in the non-perturbative relation in eq.~(\ref{Eq-41}) 
and, after all, 
indicates that 
the minimal deformation must be determined 
only by the first order perturbation 
according to the scaling behavior. 

In addition, 
we obtain 
\begin{equation}
\Gamma_{100} 
= \frac{3 \gamma_{100} t}{(t^2 + 1)^{5/2}} 
\label{Eq-43}
\end{equation}
with $\gamma_{100} = - 3 / ( 70 \sqrt{\pi} ) \approx -0.0242$ 
for the MB gas 
and 
$\gamma_{100} \approx -0.00444$ for the TF gas. 
Thus the quantum effect, i.e. difference between the MB and TF gases, 
is in the values of $\gamma_{100}$ and definitions of $G_d$. 
According to the temperature dependence in the MB gas 
($G_d \propto T^{-5/2}$), 
the deformation vanishes at the high-temperature limit; 
in other words, 
the interaction effect principally appears as the quantum effect. 

As a result, 
according to eq.~(\ref{Eq-41-2}) and $G_d > 0$, 
the interaction slightly reduces (increases) $\lambda_{p}$ 
when $\Lambda_{0} > 0$ ($< 0$). 
It is due to the angular dependence of the dipolar interaction 
in eq.~(\ref{Eq-4}). 
When $\Lambda_{0} = 0$, corresponding to the spherical trap, 
the results reproduce those of the ballistic solution, 
$\lambda = 0$, 
owing to vanishment of the interaction effect 
by the angular integration. 

Note that 
$\mathfrak{S}$ in eq.~(\ref{Eq-44}) can be given 
by the experimental setup, 
and 
$\lambda_{p}(\infty)$ ($\simeq \lambda_{0} + \lambda(\infty)$) 
can be measured from the density distribution at long-time limit 
according to eq.~(\ref{Eq-23}). 
Thus eq.~(\ref{Eq-44}) provides a theoretical approach 
to obtain the initial momentum deformation $\lambda_{0}$ 
from the experimental measurement 
as a proper correction for the TOF method. 
Because of smallness of $\gamma_{100}$ for the both gases, 
the correction is negligible 
when $\mathfrak{S}$ (or the trap anisotropy) is small; 
otherwise 
the correction must have important contribution 
to detect the interaction effects. 

Finally we should comment on difference between 
our and the previous approaches~\cite{He, Sogo} 
to treat the quadrupole deformation in the expansion. 
These are theoretically same 
except for choice of the phase-space frames: 
the expanding frame $\{ \vec{\xi}, \vec{p} \}$ in this work 
and rest frame $\{ \vec{r}, \vec{p} \}$ in the previous works. 
Thus any difference in the results must be due to 
the frame transformation. 
The previous ansatz is originally 
developed to the expanding hydrodynamical Bose gases, 
and then reproduces agreeable results 
with the experimental measurements~\cite{PS,Menotti}. 
On the other hand, 
our ansatz is made to reproduce 
the exact results for the minimal deformation 
from the ballistic solution 
in the weak interaction regime 
and give explicit description of the variation. 
The both approaches give same results 
at the ballistic limit 
for the quadratic Wigner function 
as discussed above. 
Thus we find out that 
the previous ansatz also reproduce our exact results 
up to the first order perturbation 
from the ballistic limit 
in the quadratic case. 
That is because the exact results 
depend only on the first order perturbation 
according to the scaling behavior, 
and the first order perturbation is determined 
by the ballistic solution 
according to the perturbation theory. 
This nontrivial agreement may be a significant property 
of the quadratic deformed gas 
associated with the scaling law. 

In conclusion, 
we reveal that 
the additional phase-space deformation in the expansion 
is negligible when $\mathfrak{S}$ is small, 
as shown in eq.~(\ref{Eq-44}); 
as $\mathfrak{S}$ increases, 
the deformation linearly grows 
and exhibits the scaling law associated with the Liouville's theorem 
in the long-time behaviors of the MB and TF gases. 
After all, 
our results give the exact prediction of 
the minimal quadrupole deformation 
in TDHFA and the LV approach, 
and provide 
a proper correction for the TOF method 
as a fundamental infrastructure 
for the dipolar gas physics. 
At the end, it should be noticed that 
the scaling law may not be valid 
for the large deformation by the strong interaction; 
then one should directly solve the time-evolution in the LV equation 
beyond the quadrupole ansatz 
to include the other multi-pole contribution. 
The expansion problem in the strong interaction regime should be 
studied in another paper. 
\acknowledgments
The authors would like to thank Prof. Toru Suzuki and Prof. Peter Schuck for useful discussions and comments. 
This work was supported by KAKENHI (21540412). 
%
%
%
%
%

%
%

\begin{thebibliography}{10}
\bibitem{footnote-1}
In fact, the effect on $\lambda_{p}$ 
decreases as $t^{-3}$ 
in eqs.~(\ref{Eq-41-2}) and (\ref{Eq-43}). 
It is due to the $r^{-3}$ dependence of 
the interaction.
\bibitem{footnote-3}
In principle, $\lambda_{0}$ must be determined by $G_d$ and $\Lambda_{0}$ 
in the equilibrium states; 
however we here give no relation between them 
because it has no influence on the results in this Letter.
\bibitem{footnote-5}
In general, 
the TF and MB gases 
with the one-particle self-energy in eq.~(\ref{Eq-16}) 
can be denoted 
without explicit description of $N$ and $T$ 
by using scale transformation. 
Then information of $N$ and $T$ is included 
only in the scaled parameters 
($\tilde{\vec{p}} = C \vec{p}$, 
$\tilde{\vec{r}} = C \vec{r}$, 
$\tilde{\mu}$, 
and $G_{d}$) 
depending on the statistics: 
$C = (6 N)^{- 1 / 6}$ 
and $\tilde{\mu} = (6 N)^{- 1 / 3} \mu$ 
($= 1$ in eq.~(\ref{Eq-33})) 
for the TF gas; 
$C = ( k_{\text{B}} T )^{- 1 / 2} \equiv \beta^{1 / 2} $ 
and 
$\tilde{\mu} = \ln[ N^{- 1} \beta^{-3} e^{\beta \mu} ]$ 
($= 0$ in eq.~(\ref{Eq-31})) 
for the MB gas.
\end{thebibliography}
\end{document}